\documentstyle[12pt,epsf]{article}
\newlength{\dinwidth}                                                          
\newlength{\dinmargin}                                                         
\begin{document}
\vspace{1 cm}
\newcommand{\Gev}       {\mbox{${\rm GeV}$}}
\newcommand{\Gevsq}     {\mbox{${\rm GeV}^2$}}
\newcommand{\qsd}       {\mbox{${Q^2}$}} 
\newcommand{\x}         {\mbox{${\it x}$}}
\newcommand{\smallqsd}  {\mbox{${q^2}$}} 
\newcommand{\ra}        {\mbox{$ \rightarrow $}}
\newcommand {\pom}  {I\hspace{-0.2em}P}
\newcommand {\alphapom} {\mbox{$\alpha_{_{\pom}}$}}
\newcommand {\xpom} {\mbox{$x_{_{\pom}}$}}
\newcommand {\xpomp}[1] {\mbox{$x^{#1}_{_{\pom}}\;$}}
\newcommand {\xpoma} {\mbox{$(1/x_{_{\pom}})^a\;$}}
\def\ctr#1{{\it #1}\\\vspace{10pt}}
\def\si{{\rm si}}
\def\Si{{\rm Si}}
\def\Ci{{\rm Ci}}
\def\qsq{Q^{2}}
\def\yjb{y_{_{JB}}}
\def\xjb{x_{_{JB}}}
\def\qjb{\qsq_{_{JB}}}
\def\gap{\hspace{0.5cm}}
\renewcommand{\thefootnote}{\arabic{footnote}}
\title {
\hfill{\bf \large DESY 97-047}\\
\hfill{\bf \large extended version}\\
\hfill{\bf  \large April 1997}\\
\vspace{1cm}
{\bf The Kinematics of Deep Inelastic Scattering in the Presence of Initial State Radiation
 }\\ 
\author{G\"{u}nter Wolf\\
{Deutsches Elektronen Synchrotron DESY  }\\
 }}
\date{}
\maketitle
\begin{abstract}
Initial state radiation (ISR) in deep inelastic scattering of charged leptons on nucleons can lead to substantial shifts in the reconstruction of $Q^2$ and the scaling variables $x$ and $y$. By measuring at least three out of the four variables: energy and production angle of the lepton and the hadron system, the energy of the ISR photon emitted under zero degrees can be determined. The ISR effect is illustrated for experimentation at HERA with different methods of kinematic reconstruction.    
\end{abstract}
\setcounter{page}{1}
\thispagestyle{empty}   
%
%
\section{\bf Introduction}
\label{s:intro}
The kinematics of deep-inelastic electron proton scattering (DIS) by neutral current exchange (NC),

\begin{eqnarray}
 e p \to e X, 
\label{DIS}
\end{eqnarray}

is characterized by the mass squared ($- Q^2$) of the exchanged current and the Bjorken scaling variable $x$. In a hermetic detector, $Q^2$ and $x$ can be determined from the energy and angle of the scattered electron or from the energy and momentum components of the produced hadron system or from a combination of any other two of these quantities~\cite{Bentvelsen,Hoeger,Bassler}. 

The determination of $x$ and $Q^2$ is erroneous when the incoming electron radiates a photon before the hard interaction takes place (initial state radiation, ISR) \footnote{Due to the comparatively large mass of the proton, ISR from the incoming proton can be neglected. In principle, photon emission by the scattered electron (final state radiation, FSR) will affect also the determination of the kinematics; in practice, if the scattered electron is detected in a calorimeter, the effect will be small because the signals from the electron and the FSR photon are summed.}. When averaged over a large DIS event sample, the effect of ISR in general is rather small \footnote{For DIS at HERA with $Q^2$ between 10 and 100 GeV$^2$, roughly 6$\%$ of the events have an ISR photon carrying between 5 and 20$\%$ of the energy of the incident electron.}. However, ISR can have an important effect when, due to ISR, events produced in a $Q^2 - x$ region of large cross section migrate into a region where the cross section is expected to be small. An example for such a constellation is presented by the possible excess of high-$x$, high-$Q^2$ events recently observed at HERA~\cite{H1hixhiy,Zhixhiy}. 

Most of the ISR photons are emitted very close to the direction of the electron such that the emission angle can be assumed to be zero (peaking approximation). The ISR energy and the $Q^2, x$ values can then be obtained on an event-by-event basis from a measurement of three out of the four quantities mentioned above~\cite{Bentvelsen,Hoeger,Bassler,Wulff} or from a global fit~\cite{Chaves}.

In charged current (CC) processes, 
\begin{eqnarray}
ep \to \nu X\, ,
\label{DISCC}
\end{eqnarray}
where the $\nu$ escapes detection the information from the hadron system can be used to put a limit on the energy emitted by ISR while a measurement requires direct observation of the ISR photon.

 This note presents the relevant formulae in a form suitable for HERA experiments and illustrates the migration of events due to ISR in the $x-y$, $x-Q^2$, $y-Q^2$ and $M-y$ planes, where $M$ is the effective mass of the electron current quark system.

\section{DIS kinematics in the absence of ISR}
\begin{figure}[hpbt]
\includegraphics{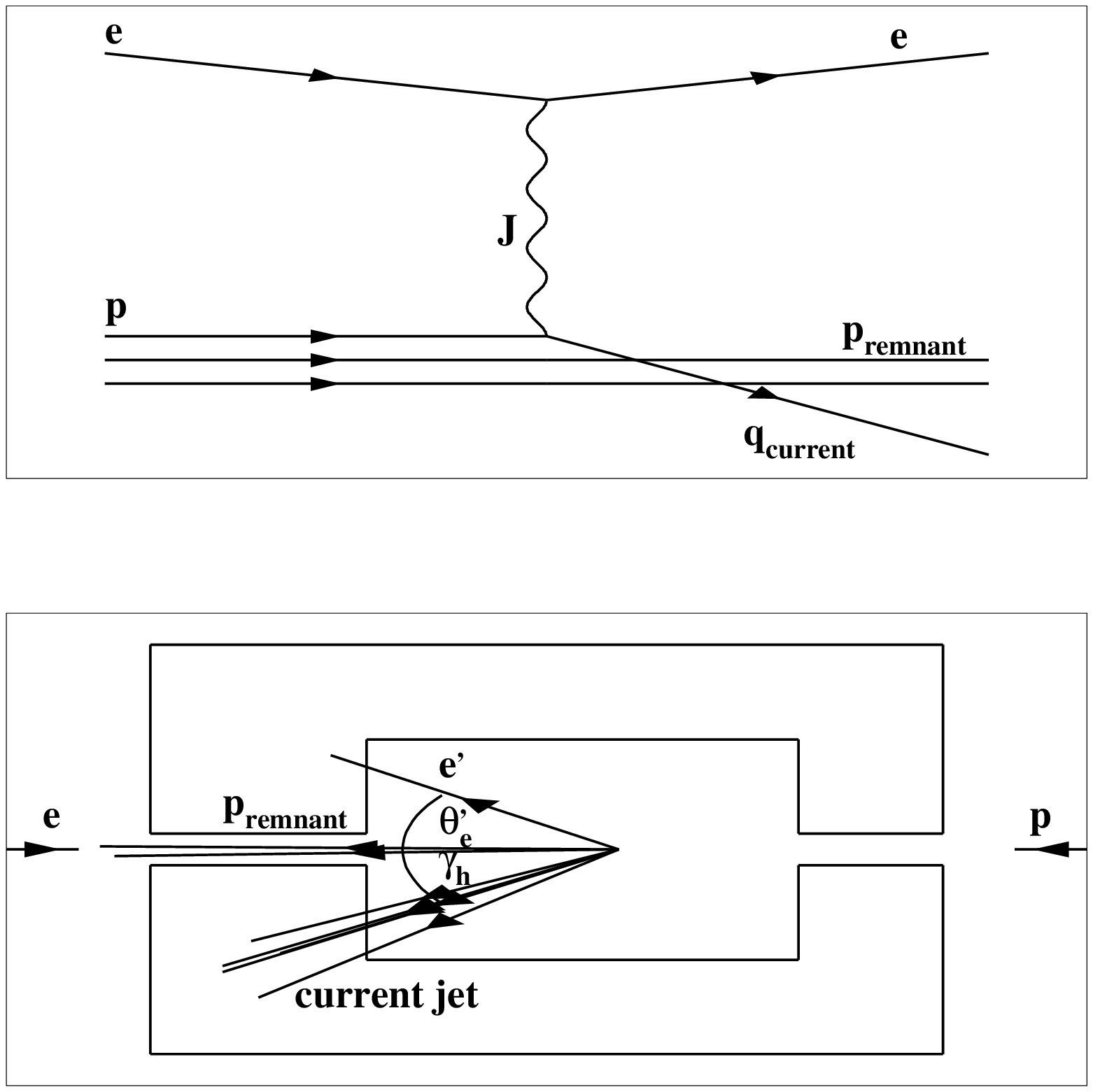}
\unitlength1cm
\begin{picture}(15,20)
\thicklines
\end{picture}
\caption{DIS, $e p \to e X$, without ISR: diagram (top) and event configuration in a HERA detector (bottom).}
\label{f:isrdiag}
\end{figure}

In deep-inelastic ep scattering (eq.~\ref{DIS}) depicted in Fig.~\ref{f:isrdiag} the incoming electron couples to the neutral electroweak current $J = \gamma, Z^0$. The relevant variables in the {\it absence} of QED radiative effects are:
\newpage
\begin{tabbing}
\hspace{0.5cm}\=$E_p$\hspace{4.8cm}                 \=proton beam energy\\
              \>$E_e$                               \> electron beam energy\\
              \>$p = (0, 0, E_p, E_p)$              \> four momentum of incoming proton with mass $m_p$\\
              \>$e = (0, 0, - E_e, E_e)$            \> four momentum of incoming electron\\
              \>$e^{\prime} = (E^{\prime}_e sin \theta^{\prime}_e, 0, E^{\prime}_e cos \theta^{\prime}_e,
                              E^{\prime}_e)$        \> four momentum of scattered electron\\ 
              \>$s = (e\,+\,p)^2 \, = \, 4 E_p E_e$ \> square of total ep c.m. energy\\
              \>$q^2 = (e\,-\,e^{\prime})^2 = -Q^2$ \> mass squared of exchanged current $J$\\
              \>                                    \> = square of four momentum transfer\\
              \>$\nu = q\cdot p/m_p$                \> energy transfer by $J$ in p rest system\\
              \>$\nu_{max} = s/(2m_p)$              \> maximum energy transfer\\          
              \>$y = (q\cdot p)/(e\cdot p) = \nu / \nu_{max}$  \> fraction of energy transfer\\
              \>$x = Q^2/(2 q\cdot p) = Q^2/(ys)$   \> Bjorken scaling variable\\
              \>$q_c = x\cdot p\,+\,(e\,-\,e^{\prime})$  \> four momentum of current quark\\
              \>$M^2 = (e^{\prime}\,+\,q_c)^2 = x\cdot s$ \> mass squared of electron - current quark system.\\
\end{tabbing} 
The electron and proton masses ($m_e, m_p$) have been neglected. The angle $\theta^{\prime}_e$ of the scattered electron is measured with respect to the incoming proton (see Fig.~\ref{f:isrdiag}b). For simplicity, the scattered electron has been assumed to have zero azimuthal angle. For fixed c.m. energy $\sqrt{s}$, the inclusive process $ep \to eX$ is described by two Lorentz-invariant variables which can be determined from the electron or the hadronic system. 

The electron side yields:\\
\begin{eqnarray}
y   & = & 1 - \frac{E^{\prime}_e}{2 E_e}(1 \, - \, cos \theta^{\prime}_e) \nonumber          \\
Q^2 & = & 2 E_e E^{\prime}_e (1 + cos \theta^{\prime}_e)                  \nonumber \\
x   & = & E^{\prime}_e(1 \, + \, cos \theta^{\prime}_e)/(2yE_p)           \nonumber \\
E^{\prime}_e           & = & (1\,-\,y)E_e \,+\,xyE_p                      \nonumber \\
cos \theta^{\prime}_e  & = & \frac{xyE_p \,-\, (1-y)E_e}{xyE_p \,+\, (1-y)E_e}   \nonumber  \\
E^{\prime \,2}_e sin^2\theta^{\prime}_e & = & 4\,xy(1-y)E_eE_p \,.       
\label{eq:emethod}
\end{eqnarray}
With the help of the relation $M\,=\,\sqrt{x\cdot s}$ the variable $x$ can be replaced by the mass $M$ of the electron current quark system.

Neglecting the proton remnant, the hadron system $j$, with energy $E_j$ and production angle $\theta_j$, yields:
\begin{eqnarray}
y   & = & \frac{E_j}{2E_e}(1\,-\,cos \theta_j)                           \nonumber            \\
Q^2 & = & E^2_j sin^2\theta_j/(1\,-\,y)                                  \nonumber   \\
x   & = & \frac{E_j}{2E_p}(1\,+\,cos \theta_j)/(1\,-\,y)                 \nonumber   \\
E_j                  & = & yE_e \,+\, x(1\,-\,y)E_p                      \nonumber   \\
cos \theta_j         & = & \frac{-\,yE_e \,+\, (1-y)xE_p}{y\,E_e \,+\, (1-y)xE_p}    \nonumber \\
E^2_j sin^2\theta_j  & = & 4\,xy(1-y)E_eE_p \, = \, Q^2\,(1-y)  \,.     
\label{eq:jmethod}
\end{eqnarray}

Using the method of Jacquet-Blondel~\cite{Jacquet} the hadron variables can be determined approximately by summing the energies ($E_h$) and transverse ($p_{Th}$) and longitudinal momenta ($p_{Zh}$) of all final state particles. The method rests on the assumption that the total transverse momentum carried by those hadrons which escape detection through the beam hole in the proton direction as well as the energy carried by particles escaping through the beam hole in the electron direction can be neglected. The result is:
\begin{eqnarray}
y_{JB}    & = & \frac{1}{2E_e}\sum_h(E_h \,-\, p_{Zh})   \nonumber                \\
Q^2_{JB}  & = & \frac{(\sum_h p_{Xh})^2 \,+\, (\sum_h p_{Yh})^2}{1 \,-\, y_{JB}}   \nonumber \\
x_{JB}    & = & Q^2_{JB}/(y_{JB}s)                      
\label{eq:JBmethod}  
\end{eqnarray}

The double-angle (DA) method of ~\cite{Bentvelsen} uses the electron scattering angle and the angle $\gamma_h$ which characterizes the longitudinal and transverse momentum flow of the hadronic system (in the naive quark-parton model $\gamma_h$ is the scattering angle of the struck quark):
\begin{eqnarray}
cos \gamma_h   =  \frac{(\sum_h p_{Xh})^2 \,+\, (\sum_h p_{Yh})^2 - (\sum_h(E_h \,-\, p_{Zh}))^2}{(\sum_h p_{Xh})^2 \,+\, (\sum_h p_{Yh})^2 + (\sum_h(E_h \,-\, p_{Zh}))^2}
\end{eqnarray}

leading to
\begin{eqnarray}
Q^2_{DA}    & = & \frac{4 E^2_e sin \gamma_h \, (1 \,+\, cos \theta^{\prime}_e)}{sin \gamma_h \,+\, sin \theta^{\prime}_e - sin(\gamma_h \,+\, \theta^{\prime}_e)}            \nonumber    \\
x_{DA}      & = & \frac{E_e}{E_p}\,\cdot\,\frac{sin \gamma_h \,+\, sin \theta^{\prime}_e + sin(\gamma_h \,+\, \theta^{\prime}_e)}{sin \gamma_h \,+\, sin \theta^{\prime}_e - sin(\gamma_h \,+\, \theta^{\prime}_e)}   
                \nonumber    \\
y_{DA}      & = & \frac{sin \gamma_h \,(1 \,+\, cos \theta^{\prime}_e)}{sin \gamma_h \,+\, sin \theta^{\prime}_e + sin(\gamma_h \,+\, \theta^{\prime}_e)}                     \nonumber  \\
            & = & \frac{(1 \,-\, cos \gamma_h) \, sin \theta^{\prime}_e}{sin \gamma_h \,+\, sin \theta^{\prime}_e - sin(\gamma_h \,+\, \theta^{\prime}_e)}                    \nonumber  \\
            & = & Q^2_{DA}/(x_{DA}s)\,. 
\label{eq:DAmethod}
\end{eqnarray}

The energy of the scattered electron obtained by the DA-method is:
\begin{eqnarray}
E^{\prime}_{eDA}   =  2\,E_e\frac{sin \gamma_h}{sin \gamma_h \,+\, sin \theta^{\prime}_e - sin(\gamma_h \,+\, \theta^{\prime}_e)}\,.
\label{eq:edavsee}       
\end{eqnarray}

The $\Sigma$ method~\cite{Bassler} combines the measurements of the energies and angles of the electron and hadron system. Making use of the relation
\begin{eqnarray}
\sum_h (E_h -p_{Zh})\,+\,E^{\prime}_e(1\,-\,cos \theta^{\prime}_e)\,=\,2E_e,
\label{eq:longitudinalmomentum}
\end{eqnarray}
which holds exactly for a hermetic detector, one can write
\begin{eqnarray}
y_{\Sigma}   & = & \frac{\sum_h(E_h -p_{Zh})}{\sum_h(E_h -p_{Zh})+E^{\prime}_e(1\,-\,cos \theta^{\prime}_e)}   \nonumber  \\
Q^2_{\Sigma} & = & \frac{E^{\prime}_e \, sin^2 \theta ^{\prime}_e}{1\,-\,y_{\Sigma}}                           \nonumber  \\
x_{\Sigma}   & = & \frac{Q^2_{\Sigma}}{s\,y_{\Sigma}}.
\label{eq:smethod}
\end{eqnarray}

\section{DIS kinematics with ISR}
\begin{figure}[hpbt]
\includegraphics{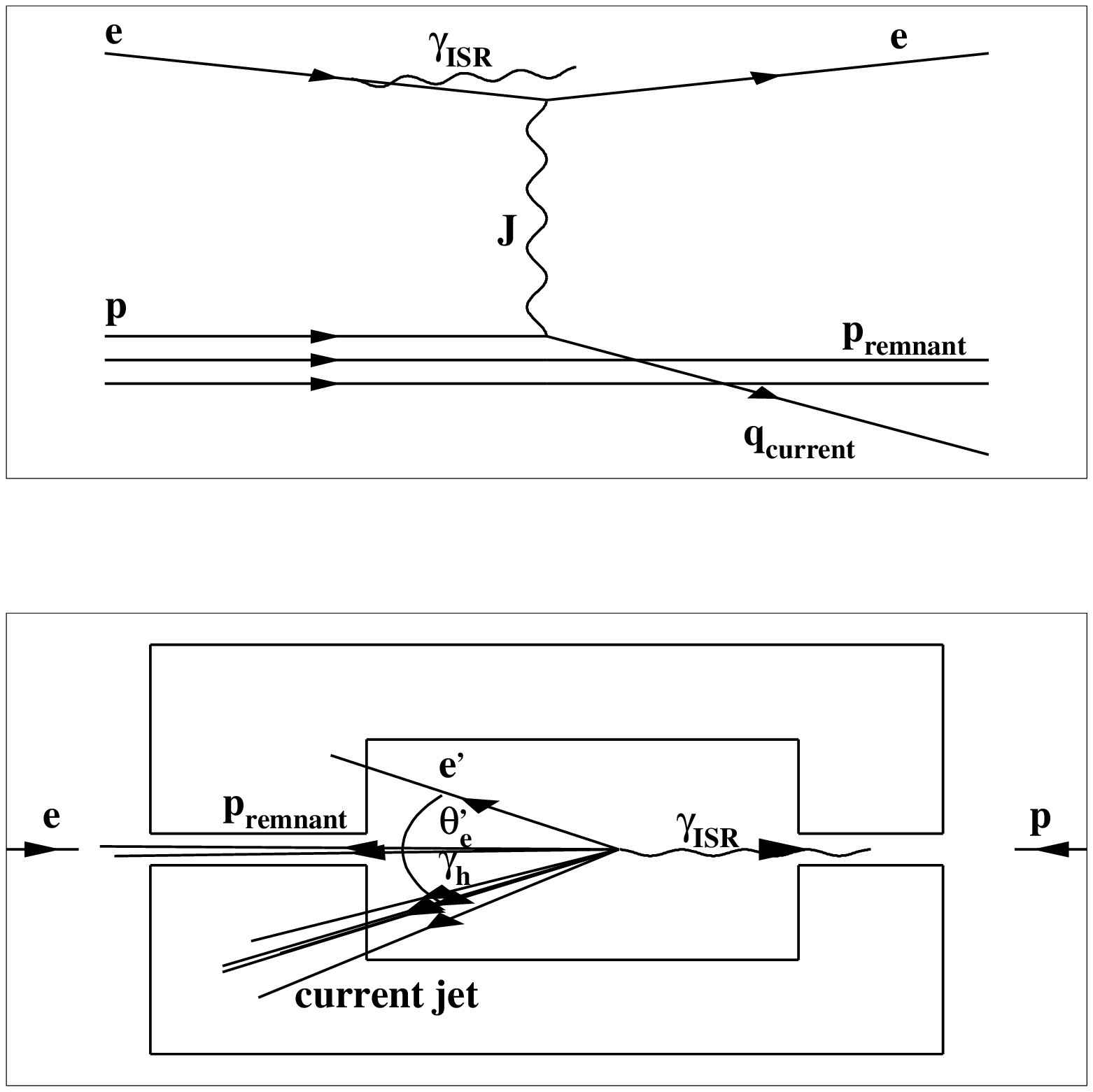}
\unitlength1cm
\begin{picture}(15,20)
\thicklines
\end{picture}
\caption{DIS, $e p \to e X$, with ISR: diagram (top) and event configuration in a HERA detector (bottom).}
\label{f:isrdiagwisr}
\end{figure}

The photon from ISR is assumed to be emitted in the direction of the beam electron (see Fig.~\ref{f:isrdiagwisr}). In order to avoid confusion, the kinematic quantities of the beam particles will be denoted by the subscript $beam$; those of the particles participating in the hard interaction will be denoted by the subscript $true$. The ``measured'' variables are those obtained under the assumption of no ISR emission. They will be identified by the method of reconstruction (i.e. e, DA or JB method) and by the subscript $meas$. They will be assumed to be measured without errors. For ease of reference all relevant variables are listed below: 

\begin{tabbing}
Initial state (nominal):\\
\\
\hspace{0.5cm}\=$E_{pbeam},\,\, E_{ebeam}$ \hspace{4.0cm}   \= proton and electron beam energies\\
              \>$s_{nominal}\, = 4 E_{pbeam}\,E_{ebeam}$    \> square of nominal ep c.m. energy\\
\\ 
ISR photon:\\
\\
              \>$E_{\gamma}$                                \> energy of the ISR photon\\
              \>$f_{\gamma}\,=\,E_{\gamma}/E_{ebeam}$       \> fraction of electron beam energy carried by ISR\\
              \>$E_{etrue}\,=\,(1\,-\, f_{\gamma}) E_{ebeam}$  \> energy of incoming electron in the hard interaction\\
\\
Hard interaction:\\
\\
              \>$x_{true},\,y_{true},\,Q^2_{true}$          \> true values\\
              \>$x_{emeas},\,y_{emeas},\,Q^2_{emeas}$       \> $x,y,Q^2$ obtained with the e-method \\
              \> $E^{\prime}_{e\,emeas},\,\theta^{\prime}_{e\,emeas}$ \> $E^{\prime}_e, \theta^{\prime}_e$ obtained with the e-method\\
              \>$x_{DAmeas},\,y_{DAmeas},\,Q^2_{DAmeas}$    \> $x, y, Q^2$ obtained with the DA method \\
              \>$E^{\prime}_{e\,DAmeas}$                      \> $E^{\prime}_{e}$ obtained with the DA-method\\
              \>$x_{JBmeas},\,y_{JBmeas},\,Q^2_{JBmeas}$    \> $x,y,Q^2$ obtained with the JB method\\ 
              \>$x_{\Sigma},\,y_{\Sigma},\,Q^2_{\Sigma}$    \> $x,y,Q^2$ obtained with the $\Sigma$ method
\end{tabbing}

The effect of ISR is taken into account by replacing in eqs.~\ref{eq:emethod} to~\ref{eq:DAmethod} the electron beam energy by $E_{etrue}$. The result is
\begin{eqnarray}
s_{true}     & = & (1\,-\,f_{\gamma}) s_{nominal}\,
\end{eqnarray}
and for the electron method:
\begin{eqnarray}
x_{true}     & = & x_{emeas} \frac{y_{emeas}}{y_{true}} \\
             & = & x_{emeas}\frac{y_{emeas}(1\,-\,f_{\gamma})}{y_{emeas}\,-\,f_{\gamma}}  \nonumber  \\
y_{true}     & = & \frac{y_{emeas}\,-\,f_{\gamma}}{1\,-\,f_{\gamma}}  \nonumber   \\
Q^2_{true}   & = & Q^2_{emeas} (1\,-\,f_{\gamma})\,.  \nonumber
\label{emethodisr}
\end{eqnarray} 

One finds for the electron method that
\begin{itemize}
\item $f_{\gamma}$ cannot exceed $y_{emeas}$: $f_{\gamma} < y_{emeas}$. $f_{\gamma}$ is limited also by the condition $x_{true} \le 1$ which leads to $f_{\gamma} \le \frac{y_{emeas}(1-x_{emeas})}{1-x_{emeas}y_{emeas}}$;
\item all three variables ($x, y, Q^2$) are affected by ISR. In the presence of ISR the points move in the $x,y$ plane along lines of constant $Q^2_{emeas}$: note, $x_{emeas}y_{emeas} = x_{true}y_{true}$ and therefore $Q^2_{emeas} = s_{nominal} x_{emeas}y_{emeas} = s_{nominal} x_{true}y_{true}$. Since ISR reduces $s$ the true $Q^2$ value is smaller than $Q^2_{emeas}$.
\item the measured $x$ value is smaller than the true one, $x_{emeas} \,<\, x_{true}$;
\item the measured $y$ value is larger than the true one, $y_{emeas} \,>\,y_{true}$. At small $y$ the shift in $y$ - and therefore also in $x$ - can become quite large, e.g. for $f_{\gamma} = 0.1$, $y_{true} = 0.2$ one finds $y_{emeas} = 1.4 y_{true}$ and $x_{emeas} = 0.7 x_{true}$. 

\end{itemize}

The corresponding results for the DA variables are:
\begin{eqnarray}
x_{true}    & = & (1\,-\,f_{\gamma})\cdot x_{DAmeas}                     \\
y_{true}    & = & y_{DAmeas}                                  \nonumber  \\
Q^2_{true}  & = & Q^2_{DAmeas} \cdot (1\,-\,f_{\gamma})^2\,.    \nonumber  
\label{DAmethodisr}
\end{eqnarray}
 
One finds for the DA method that
\begin{itemize}
\item $x$ and $Q^2$ are affected by ISR while $y$ is insensitive to ISR;
\item the measured $x$ value is larger than the true one, $x_{DAmeas} \,>\, x_{true}$;
\item ISR has a strong effect on $Q^2$ since the factor $(1\,-\,f_{\gamma})$ enters quadratically. To give an example, for $f_{\gamma} = 0.2$ and $Q^2_{true} = 20 000$ GeV$^2$, the measured $Q^2$ value is more than $50 \%$ larger, viz. $Q^2_{DAmeas} = 31250$ GeV$^2$.
\end{itemize}

The results for the Jacquet-Blondel method are:
\begin{eqnarray}
x_{true}   & = & x_{JBmeas} \frac{1\,-\,y_{JBmeas}}{1\,-\,y_{JBmeas}/(1\,-\,f_{\gamma})}                            \\
y_{true}   & = & y_{JBmeas} \frac{1}{1\,-\,f_{\gamma}}   \nonumber \\
Q^2_{true} & = & Q^2_{JBmeas} \frac{1\,-\,y_{JBmeas}}{1\,-\,y_{JBmeas}/(1\,-\,f_{\gamma})}                                                   \nonumber  
\label{JBmethodisr}
\end{eqnarray}
One finds for the JB method that
\begin{itemize}
\item $f_{\gamma}$ cannot exceed $(1\,-\,y_{JBmeas})$: $f_{\gamma} < (1\,-\,y_{JBmeas}$). $f_{\gamma}$ is limited also by the condition $x_{true} \le 1$ which leads to $f_{\gamma} \le \frac{(1-x_{JBmeas})(1-y_{JBmeas})}{1-x_{JBmeas}(1-y_{JBmeas})}$;
\item $x,y$ and $Q^2$ are affected by ISR.
\end{itemize}

The results for the $\Sigma$ method are:
\begin{eqnarray}
x_{true}   & = & \frac{x_{\Sigma meas}}{1\,-\,f_{\gamma}} \\
y_{true}   & = & y_{\Sigma}                                \nonumber  \\
Q^2_{true} & = & Q^2_{\Sigma}.                             \nonumber
\end{eqnarray} 
One finds for the $\Sigma$ method that
\begin{itemize}
\item $y$ and $Q^2$ are not affected by ISR. This is so because eq.~\ref{eq:longitudinalmomentum} holds also in the presence of ISR when $E_{etrue}$ is used for $E_e$ instead of $E_{ebeam}$.
\item Note that $M$ is also unaffected by ISR since $M^2 = x_{true}s_{true} = x_{\Sigma meas}s_{nominal}$.
\end{itemize}

\subsection{Graphical illustration of ISR effects}
\begin{figure}[hpbt]
\includegraphics{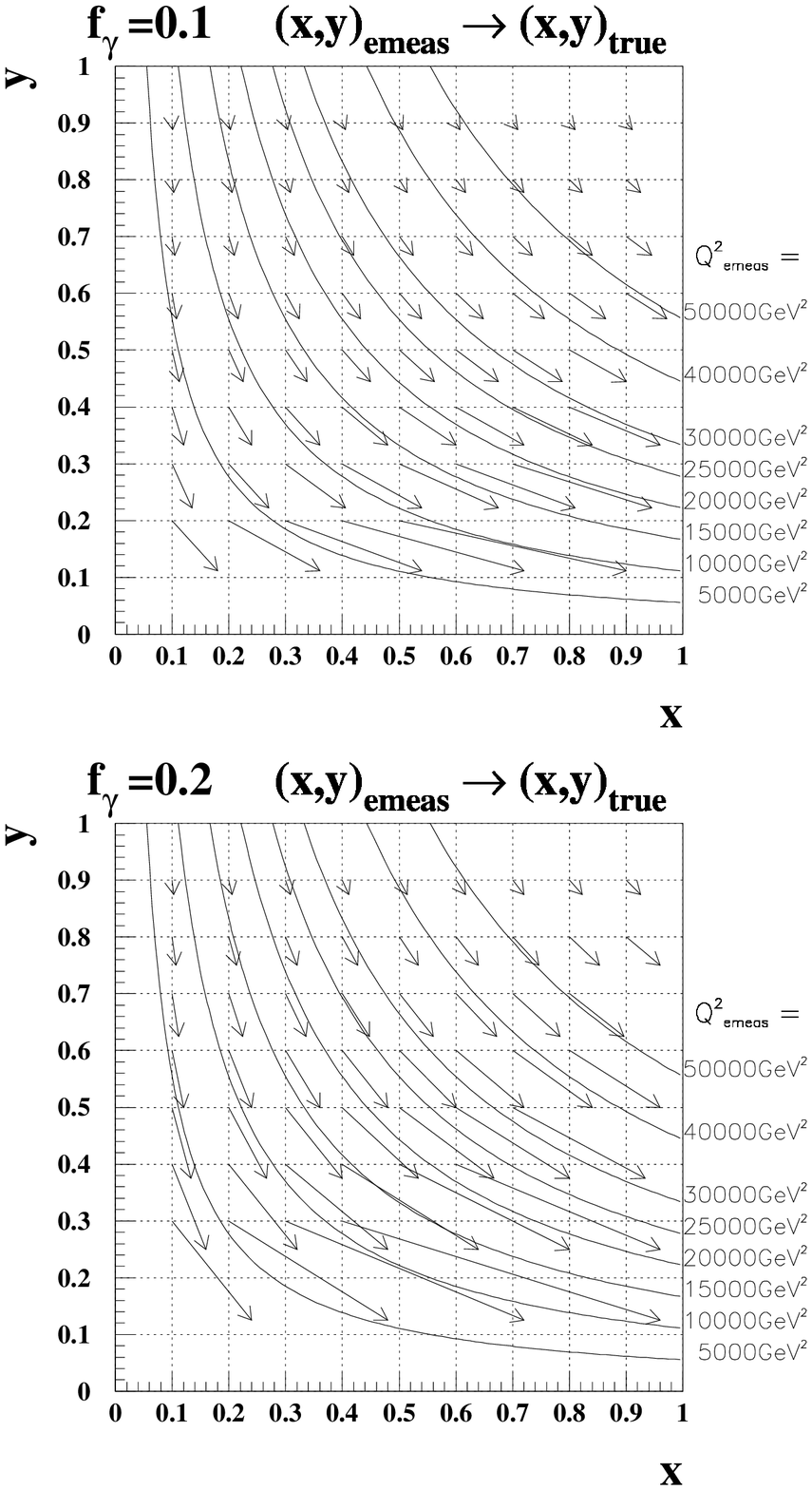}
\unitlength1cm
\begin{picture}(15,18)
\thicklines
\end{picture}
\caption{The $x\,-\,y$ plane: The effect of ISR on the determination of $x$ and $y$ by the electron method for $f_{\gamma} \equiv \frac{E_{\gamma}}{E_{ebeam}} = 0.1$ (top) and 0.2 (bottom). The arrows point from the point $(x_{emeas},y_{emeas})$ reconstructed with the electron method to the point $(x_{true},y_{true})$. The curves show lines with constant $Q^2_{emeas}$.}
\label{f:isrxye}
\end{figure}

\begin{figure}[hpbt]
\includegraphics{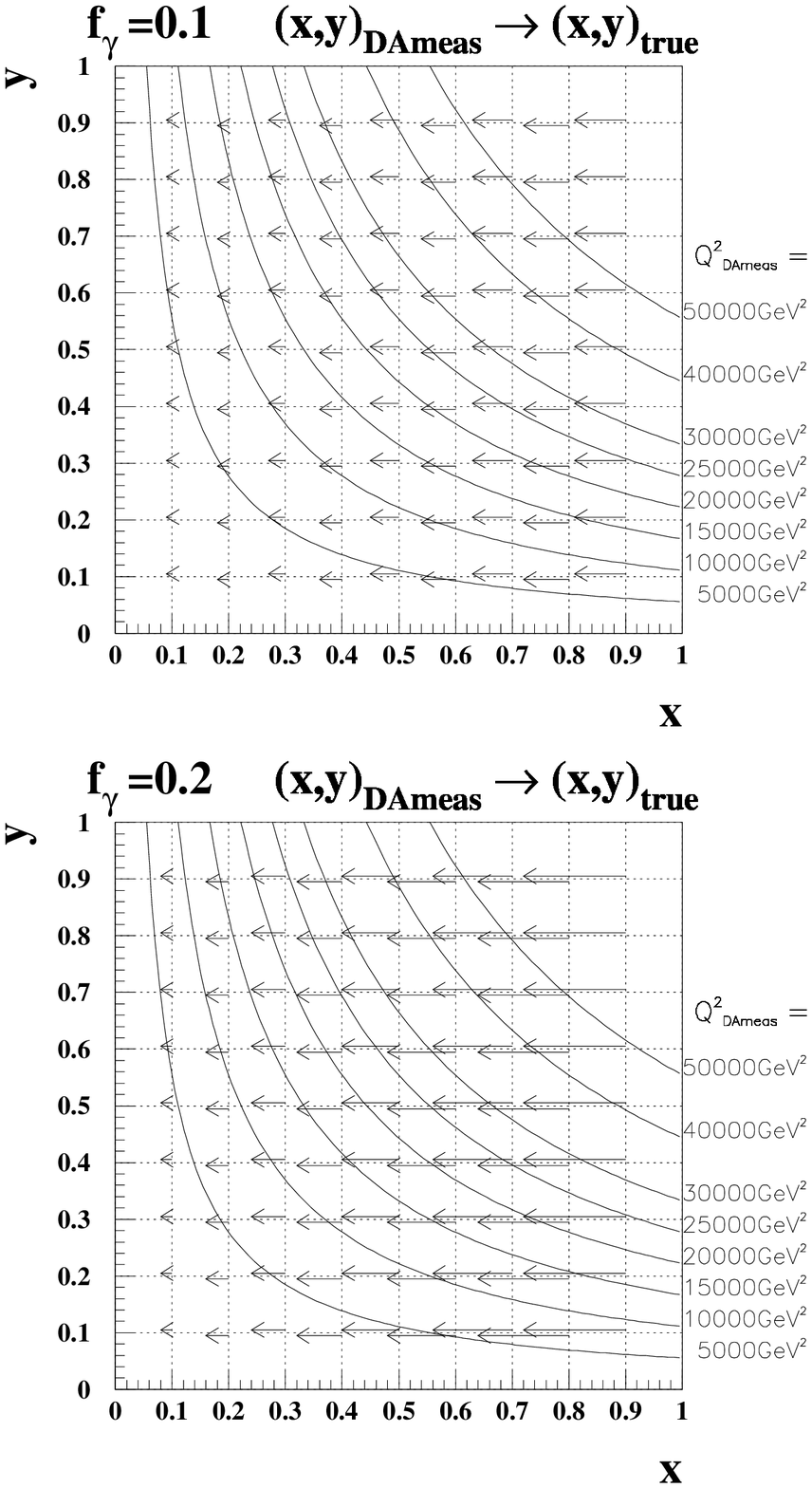}
\unitlength1cm
\begin{picture}(15,18)
\thicklines
\end{picture}
\caption{The $x\,-\,y$ plane: The effect of ISR on the determination of $x$ and $y$ by the DA method for $f_{\gamma} \equiv \frac{E_{\gamma}}{E_{ebeam}} = 0.1$ (top) and 0.2 (bottom).  The arrows point from the point $(x_{DAmeas},y_{DAmeas})$ reconstructed with the DA method to the point $(x_{true},y_{true})$. The curves show lines with constant $Q^2_{DAmeas}$.}
\label{f:isrxyda}
\end{figure}

\begin{figure}[hpbt]
\includegraphics{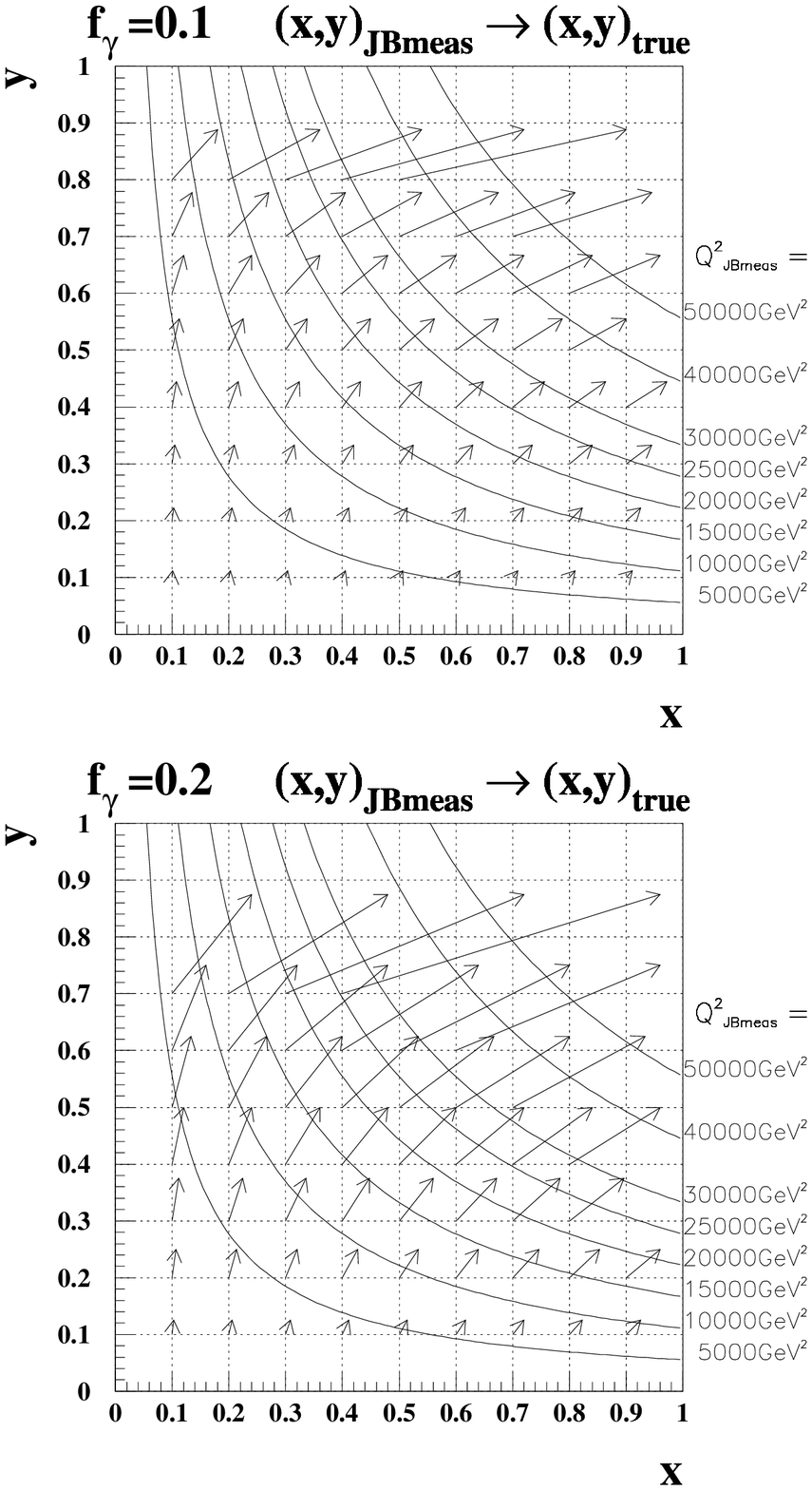}
\unitlength1cm
\begin{picture}(15,18)
\thicklines
\end{picture}
\caption{The $x\,-\,y$ plane: The effect of ISR on the determination of $x$ and $y$ by the JB method for $f_{\gamma} \equiv \frac{E_{\gamma}}{E_{ebeam}} = 0.1$ (top) and 0.2 (bottom).  The arrows point from the point $(x_{JBmeas},y_{JBmeas})$ reconstructed with the JB method to the point $(x_{true},y_{true})$. The curves show lines with constant $Q^2_{JBmeas}$.}
\label{f:isrxyjb}
\end{figure}

\begin{figure}[hpbt]
\includegraphics{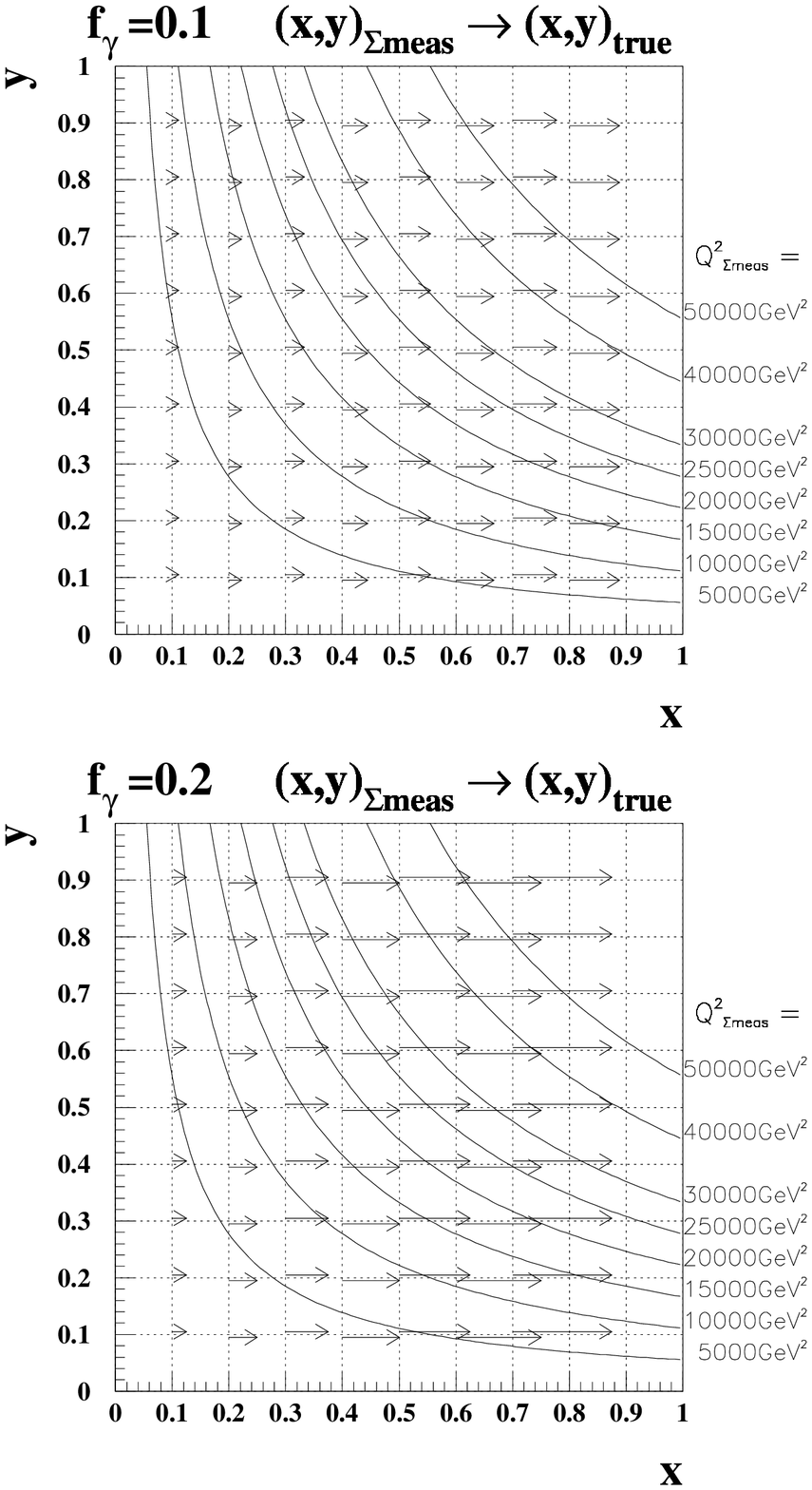}
\unitlength1cm
\begin{picture}(15,18)
\thicklines
\end{picture}
\caption{The $x\,-\,y$ plane: The effect of ISR on the determination of $x$ and $y$ by the $\Sigma$ method for $f_{\gamma} \equiv \frac{E_{\gamma}}{E_{ebeam}} = 0.1$ (top) and 0.2 (bottom). The arrows point from the point $(x_{\Sigma meas},y_{\Sigma meas})$ reconstructed with the $\Sigma$ method to the point $(x_{true},y_{true})$. The curves show lines with constant $Q^2_{\Sigma meas}$.}
\label{f:isrxys}
\end{figure}

\begin{figure}[hpbt]
\includegraphics{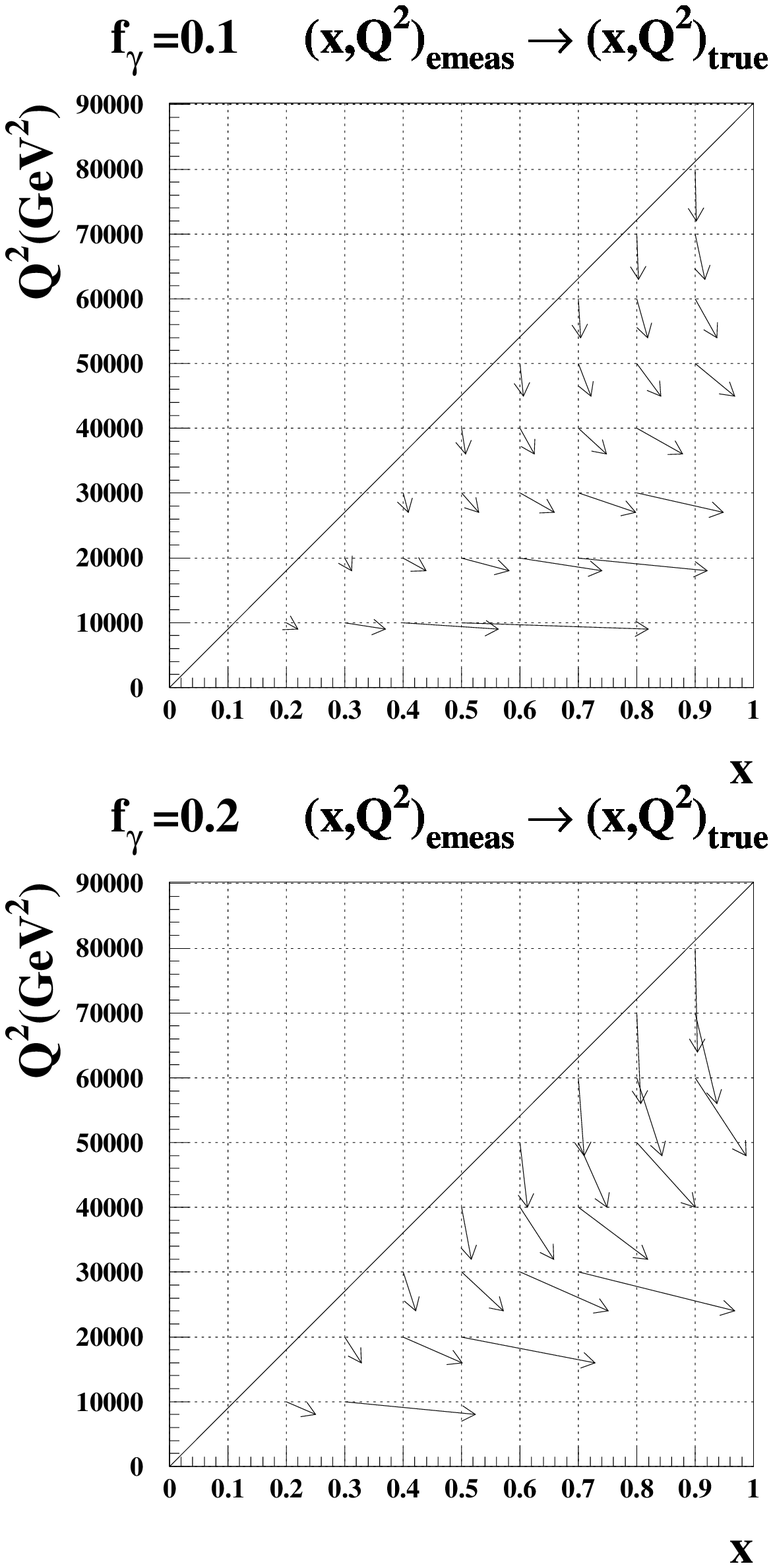}
\unitlength1cm
\begin{picture}(15,18)
\thicklines
\end{picture}
\caption{The $x\,-\,Q^2$ plane: The effect of ISR on the determination of $x$ and $Q^2$ by the electron method for $f_{\gamma} \equiv \frac{E_{\gamma}}{E_{ebeam}} = 0.1$ (top) and 0.2 (bottom).  The arrows point from the point $(x_{emeas},Q^2_{emeas})$ reconstructed with the electron method to the point $(x_{true},Q^2_{true})$.}
\label{f:isrxqe}
\end{figure}

\begin{figure}[hpbt]
\includegraphics{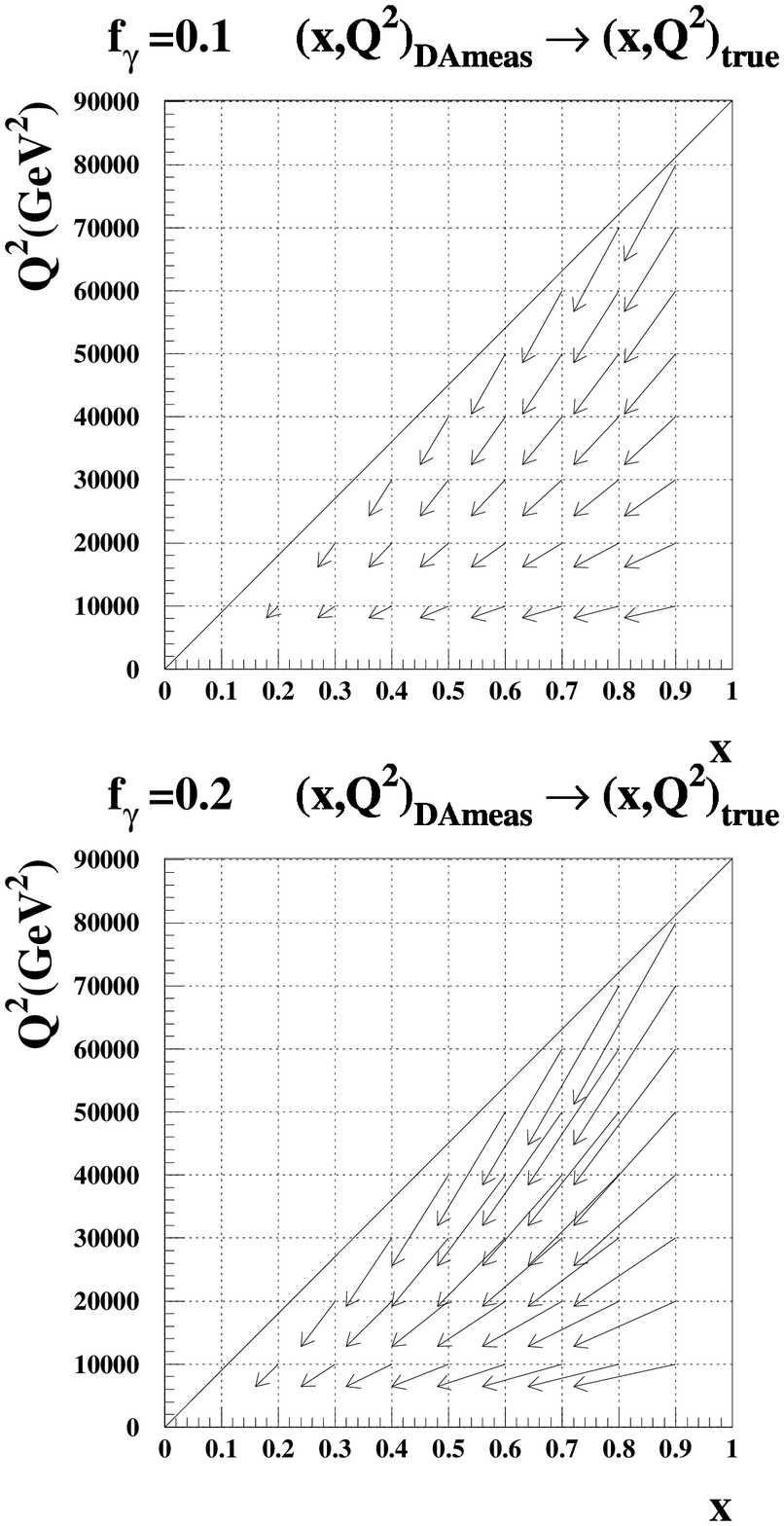}
\unitlength1cm
\begin{picture}(15,18)
\thicklines
\end{picture}
\caption{The $x\,-\,Q^2$ plane: The effect of ISR on the determination of $x$ and $Q^2$ by the DA method for $f_{\gamma} \equiv \frac{E_{\gamma}}{E_{ebeam}} = 0.1$ (top) and 0.2 (bottom).  The arrows point from the point $(x_{DAmeas},Q^2_{DAmeas})$ reconstructed with the DA method to the point $(x_{true},Q^2_{true})$.}
\label{f:isrxqda}
\end{figure}

\begin{figure}[hpbt]
\includegraphics{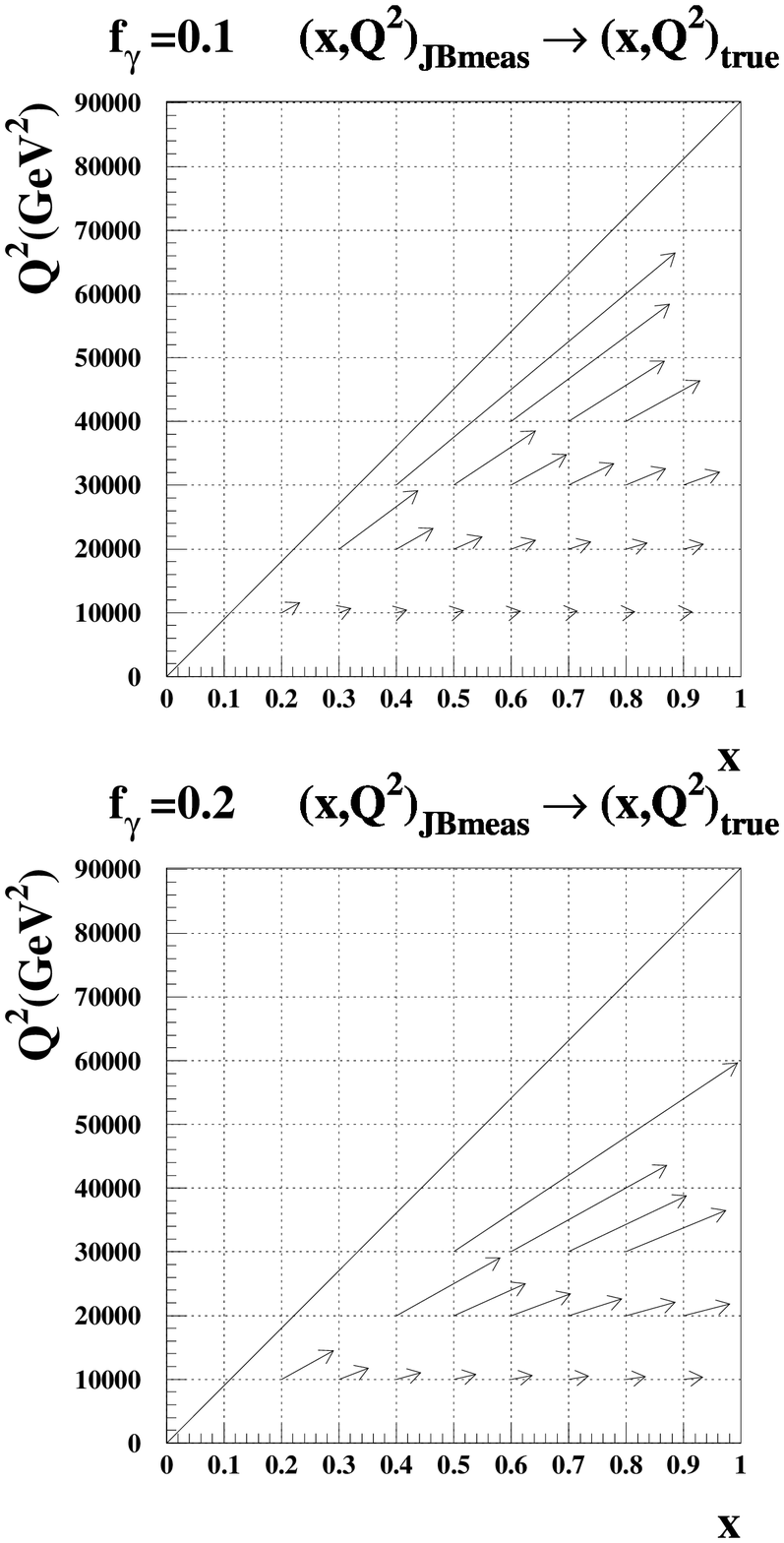}
\unitlength1cm
\begin{picture}(15,18)
\thicklines
\end{picture}
\caption{The $x\,-\,Q^2$ plane: The effect of ISR on the determination of $x$ and $Q^2$ by the JB method for $f_{\gamma} \equiv \frac{E_{\gamma}}{E_{ebeam}} = 0.1$ (top) and 0.2 (bottom).  The arrows point from the point $(x_{JBmeas},Q^2_{JBmeas})$ reconstructed with the JB method to the point $(x_{true},Q^2_{true})$.}
\label{f:isrxqjb}
\end{figure}

\begin{figure}[hpbt]
\includegraphics{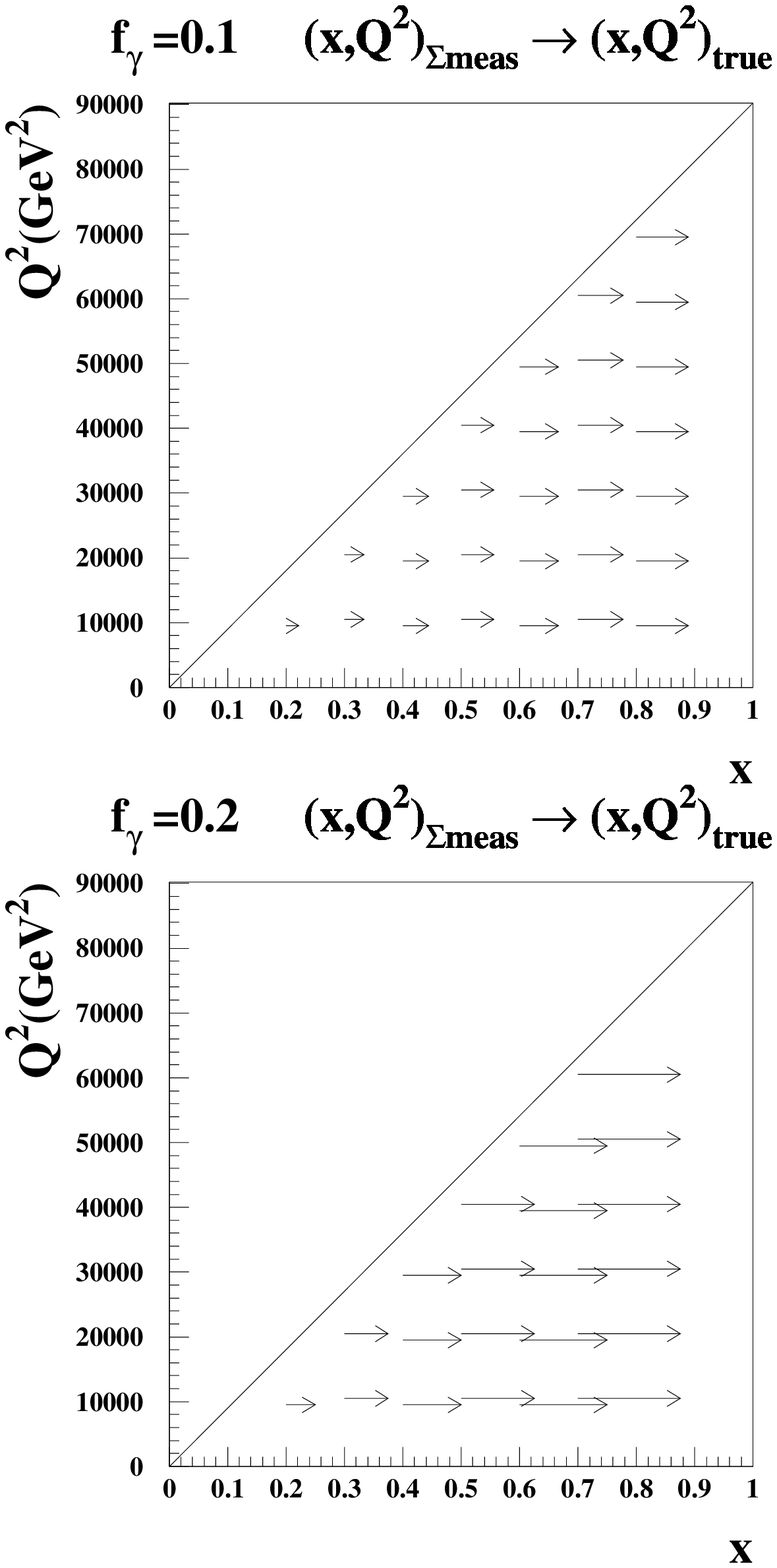}
\unitlength1cm
\begin{picture}(15,18)
\thicklines
\end{picture}
\caption{The $x\,-\,Q^2$ plane: The effect of ISR on the determination of $x$ and $Q^2$ by the $\Sigma$ method for $f_{\gamma} \equiv \frac{E_{\gamma}}{E_{ebeam}} = 0.1$ (top) and 0.2 (bottom).  The arrows point from the point $(x_{\Sigma meas},Q^2_{\Sigma meas})$ reconstructed with the $\Sigma$ method to the point $(x_{true},Q^2_{true})$.}
\label{f:isrxqs}
\end{figure}

\begin{figure}[hpbt]
\includegraphics{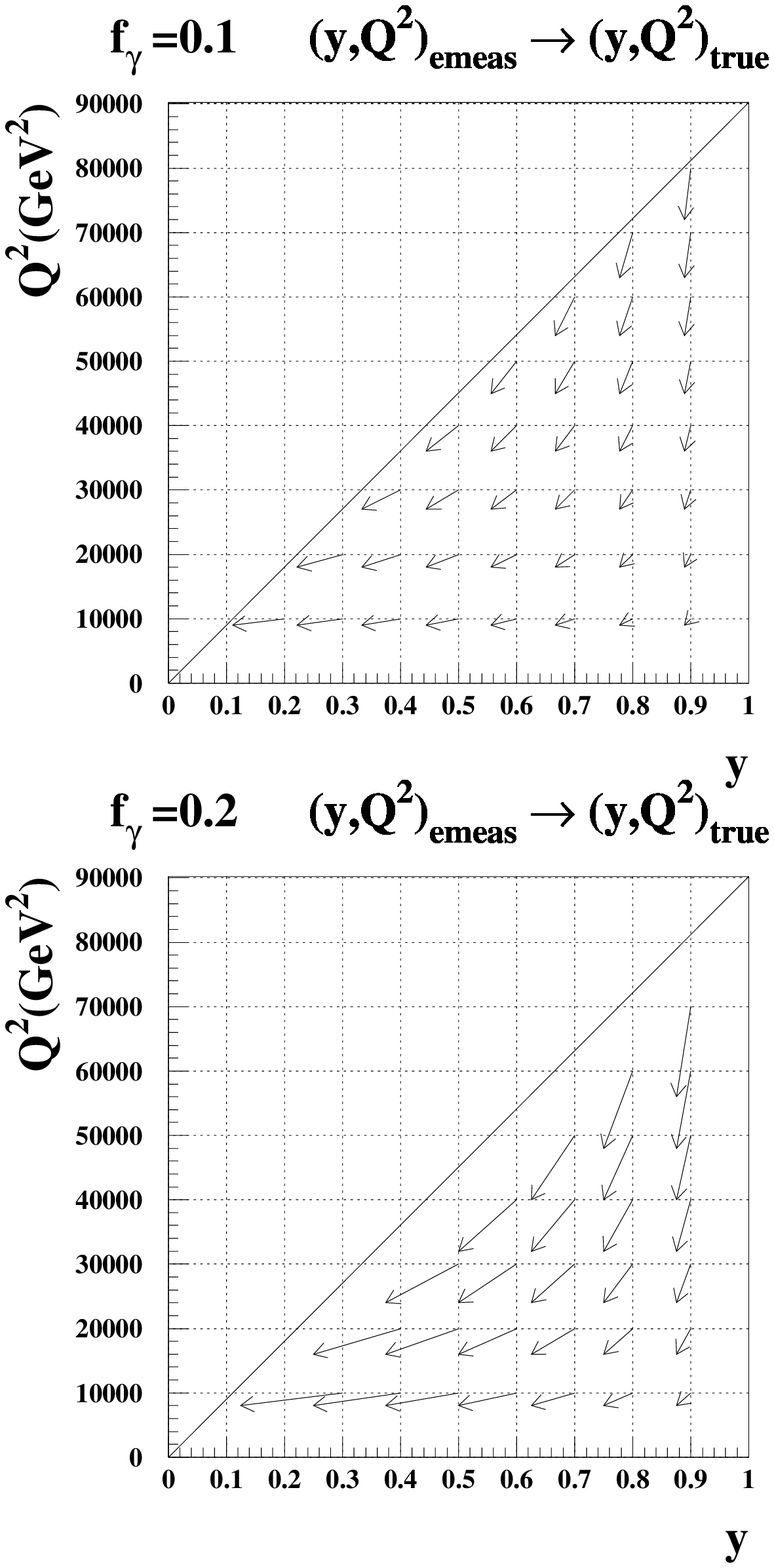}
\unitlength1cm
\begin{picture}(15,18)
\thicklines
\end{picture}
\caption{The $y\,-\,Q^2$ plane: The effect of ISR on the determination of $y$ and $Q^2$ by the electron method for $f_{\gamma} \equiv \frac{E_{\gamma}}{E_{ebeam}} = 0.1$ (top) and 0.2 (bottom).  The arrows point from the point $(y_{emeas},Q^2_{emeas})$ reconstructed with the electron method to the point $(y_{true},Q^2_{true})$.}
\label{f:isryqe}
\end{figure}

\begin{figure}[hpbt]
\includegraphics{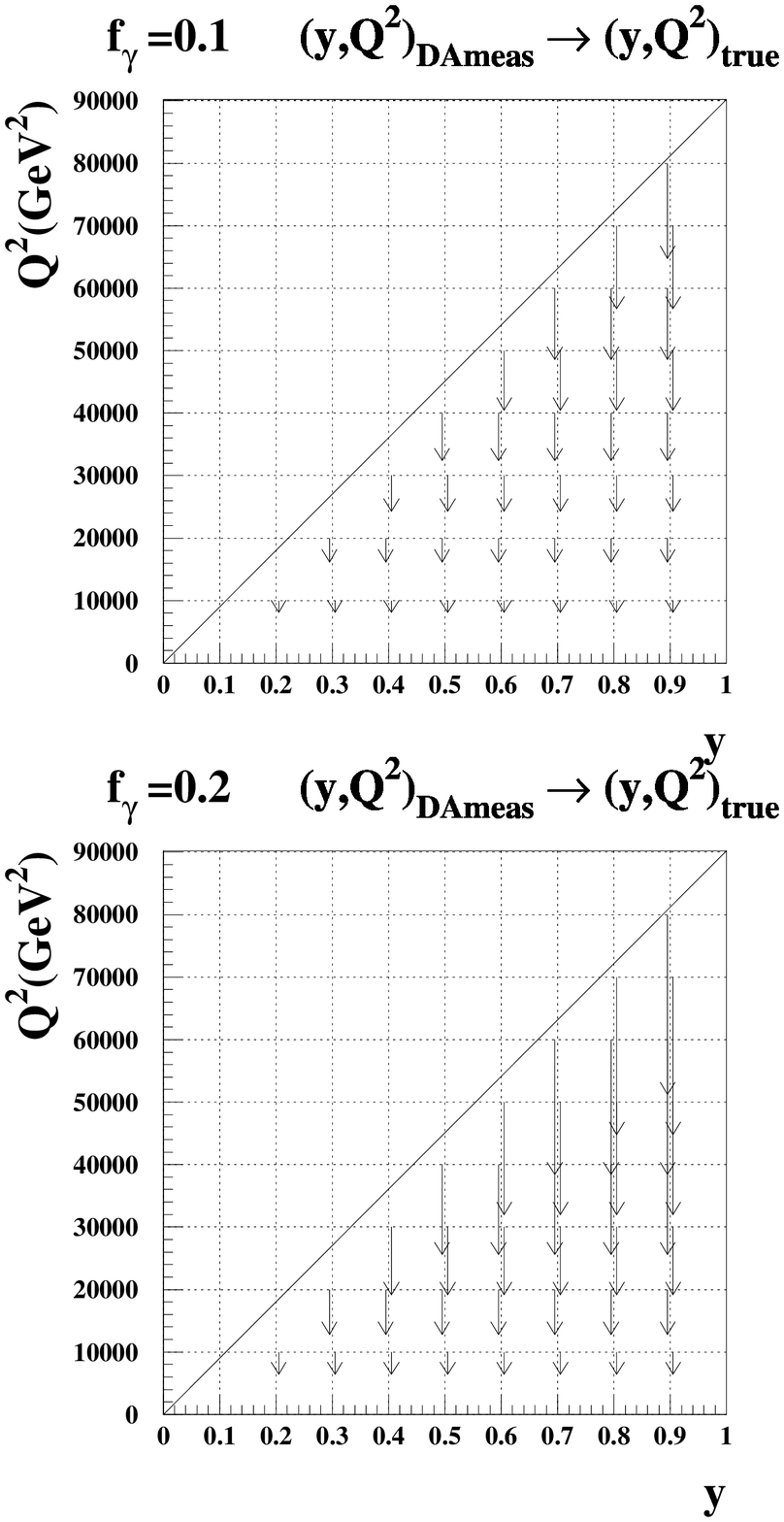}
\unitlength1cm
\begin{picture}(15,18)
\thicklines
\end{picture}
\caption{The $y\,-\,Q^2$ plane: The effect of ISR on the determination of $y$ and $Q^2$ by the DA method for $f_{\gamma} \equiv \frac{E_{\gamma}}{E_{ebeam}} = 0.1$ (top) and 0.2 (bottom).  The arrows point from the point $(y_{DAmeas},Q^2_{DAmeas})$ reconstructed with the DA method to the point $(y_{true},Q^2_{true})$.}
\label{f:isryqda}
\end{figure}

\begin{figure}[hpbt]
\includegraphics{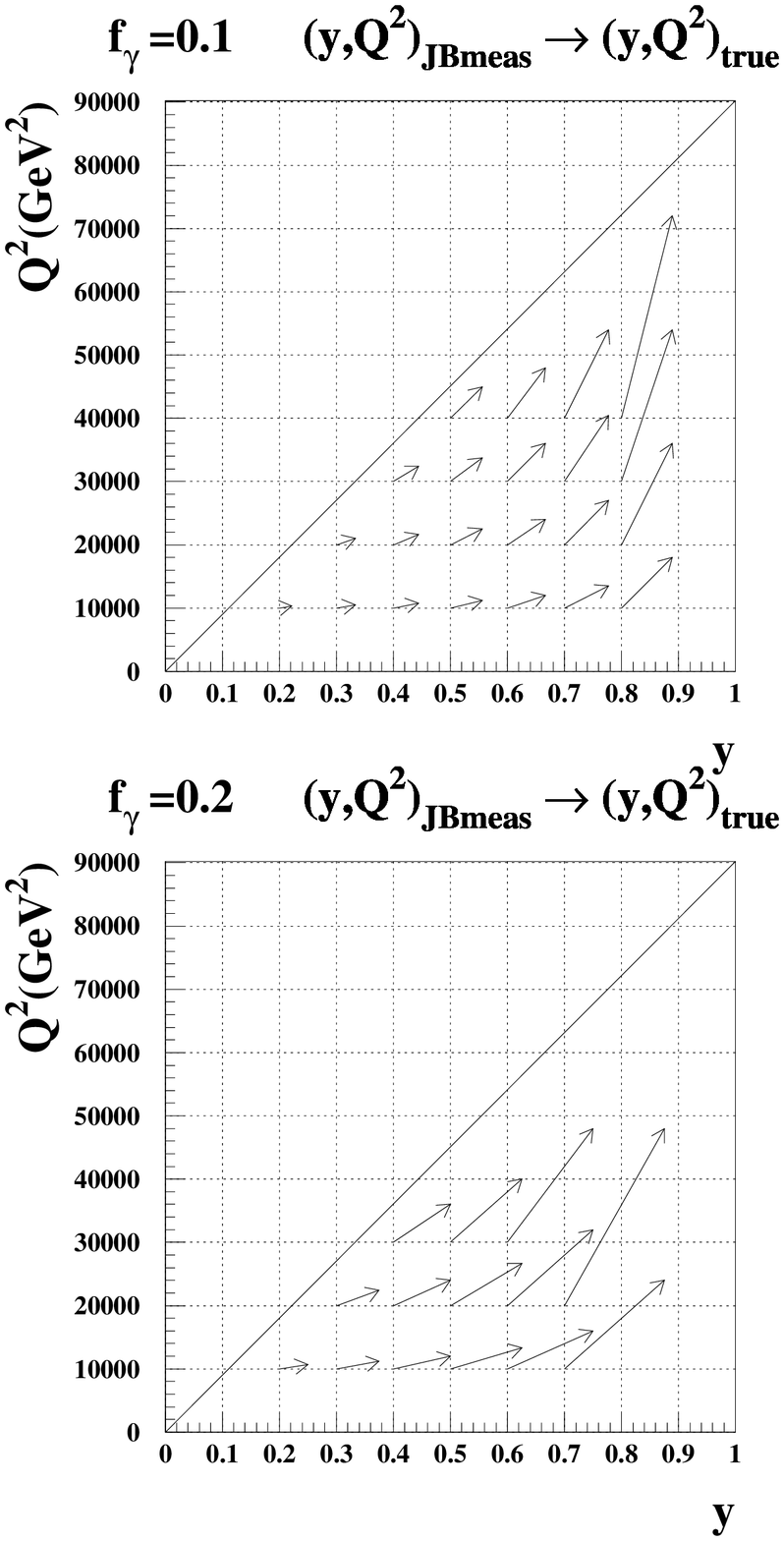}
\unitlength1cm
\begin{picture}(15,18)
\thicklines
\end{picture}
\caption{The $y\,-\,Q^2$ plane: The effect of ISR on the determination of $y$ and $Q^2$ by the JB method for $f_{\gamma} \equiv \frac{E_{\gamma}}{E_{ebeam}} = 0.1$ (top) and 0.2 (bottom). The arrows point from the point $(y_{JBmeas},Q^2_{JBmeas})$ reconstructed with the JB method to the point $(y_{true},Q^2_{true})$.}
\label{f:isryqjb}
\end{figure}

\begin{figure}[hpbt]
\includegraphics{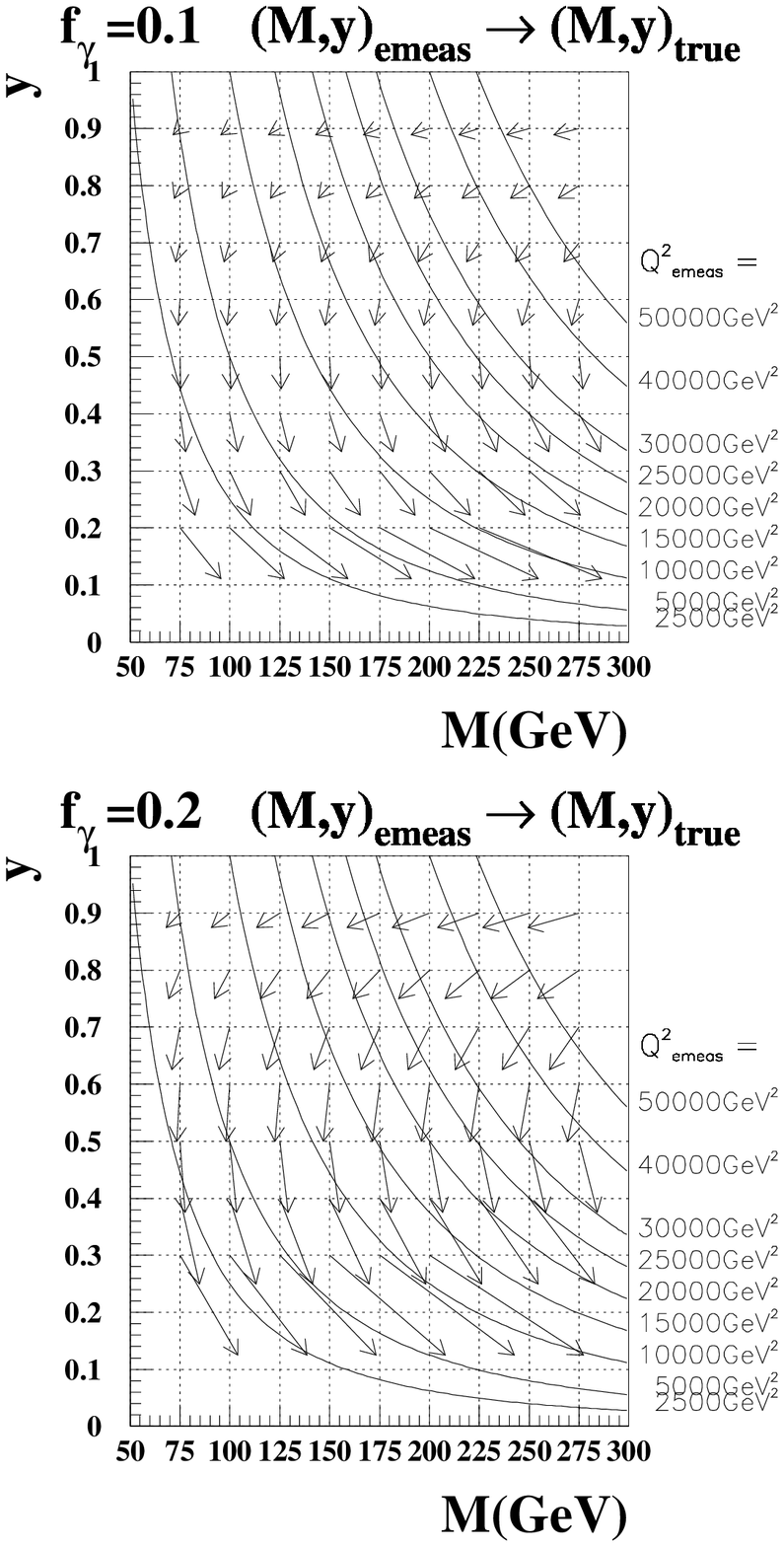}
\unitlength1cm
\begin{picture}(15,18)
\thicklines
\end{picture}
\caption{The $M\,-\,y$ plane: The effect of ISR on the determination of $M = \sqrt{xs}$ and $y$ by the electron method for $f_{\gamma} \equiv \frac{E_{\gamma}}{E_{ebeam}} = 0.1$ (top) and 0.2 (bottom).  The arrows point from the point $(M_{emeas},y_{emeas})$ reconstructed with the electron method to the point $(M_{true},y_{true})$. The curves show lines with constant $Q^2_{emeas}$.}
\label{f:isrmye}
\end{figure}

\begin{figure}[hpbt]
\includegraphics{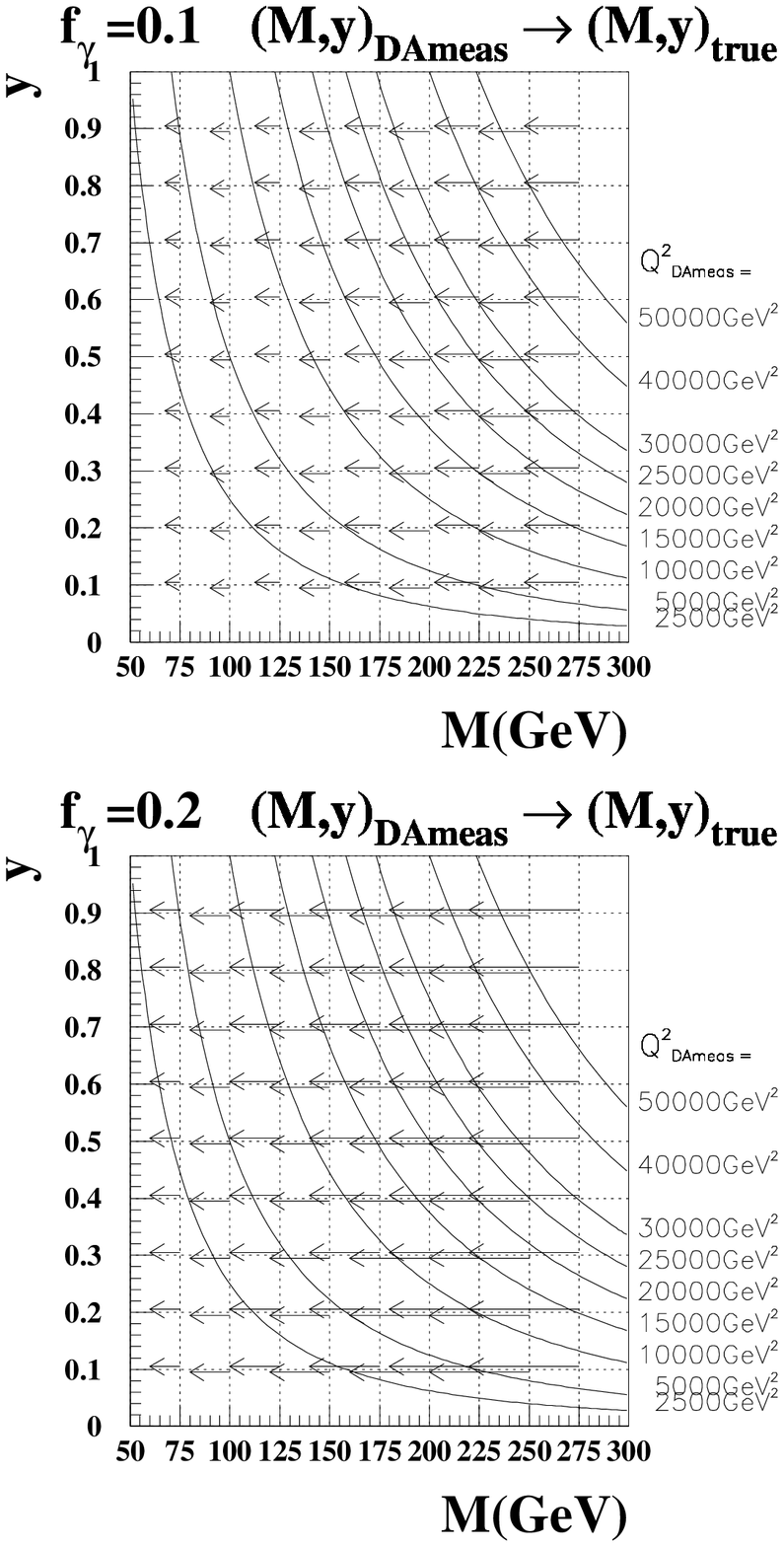}
\unitlength1cm
\begin{picture}(15,18)
\thicklines
\end{picture}
\caption{The $M\,-\,y$ plane: The effect of ISR on the determination of $M = \sqrt{xs}$ and $y$ by the DA method for $f_{\gamma} \equiv \frac{E_{\gamma}}{E_{ebeam}} = 0.1$ (top) and 0.2 (bottom).  The arrows point from the point $(M_{DAmeas},y_{DAmeas})$ reconstructed with the DA method to the point $(M_{true},y_{true})$. The curves show lines with constant $Q^2_{DAmeas}$.}
\label{f:isrmyda}
\end{figure}

\begin{figure}[hpbt]
\includegraphics{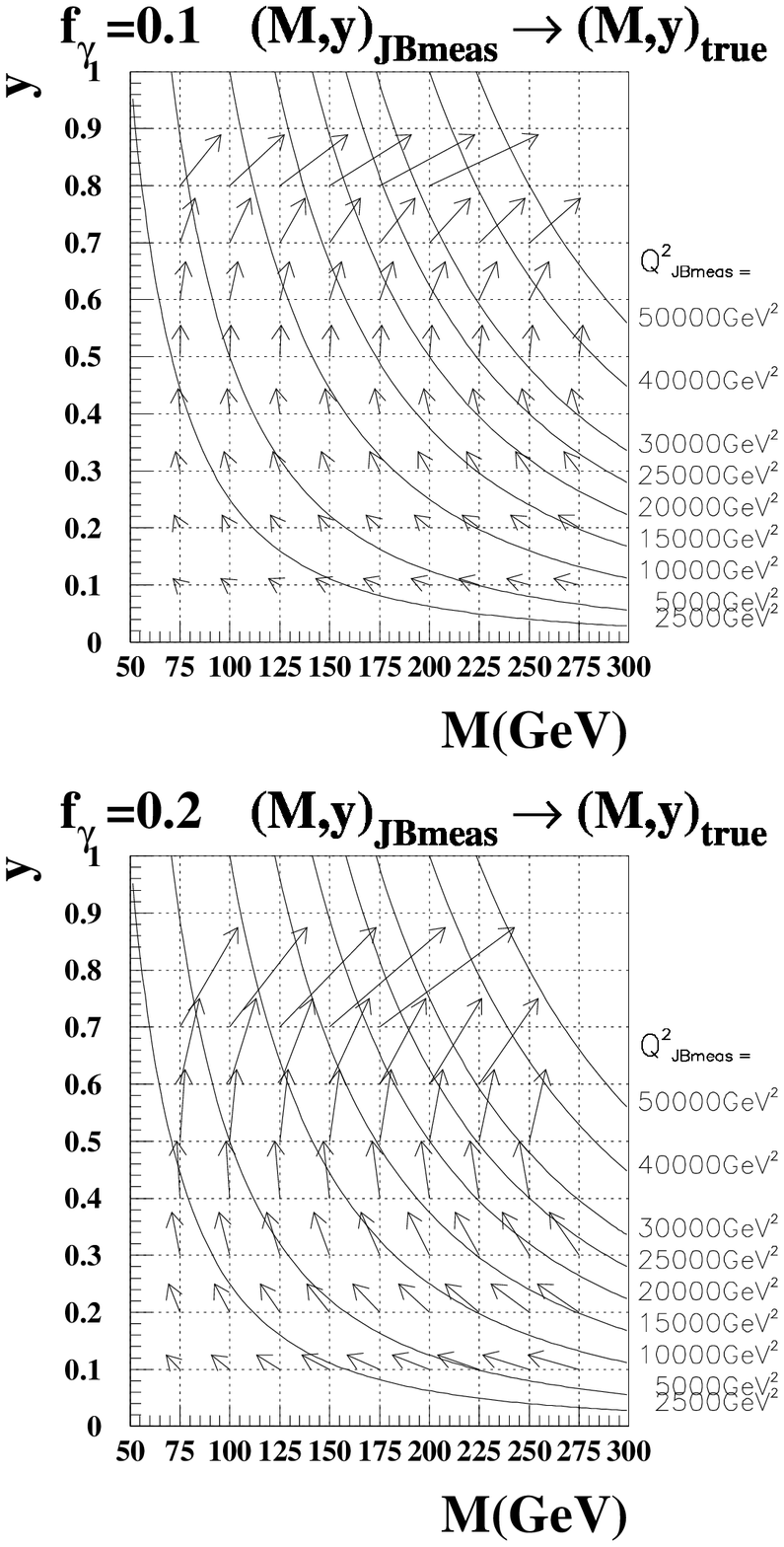}
\unitlength1cm
\begin{picture}(15,18)
\thicklines
\end{picture}
\caption{The $M\,-\,y$ plane: The effect of ISR on the determination of $M = \sqrt{xs}$ and $y$ by the JB method for $f_{\gamma} \equiv \frac{E_{\gamma}}{E_{ebeam}} = 0.1$ (top) and 0.2 (bottom).  The arrows point from the point $(M_{JBmeas},y_{JBmeas})$ reconstructed with the JB method to the point $(M_{true},y_{true})$. The curves show lines with constant $Q^2_{JBmeas}$.}
\label{f:isrmyjb}
\end{figure}

The effects of ISR on the measurements of $x,y$ and $Q^2$ when using the e, DA, JB and $\Sigma$ methods have been calculated for collisions of $E_{ebeam} = 27.5$ GeV electrons on $E_{pbeam} = 820$ GeV protons. The results for the $x\,-\,y$,  $x\,-\,Q^2$, $y\,-\,Q^2$ and $M\,-\,y$ planes are presented in 
Figs.~\ref{f:isrxye} to ~\ref{f:isrmyjb} 
for two values of the fractional energy carried away by the radiated photon: $f_{\gamma} = 0.1$ and 0.2. The arrows shown point from the measured to the true quantities. The curves in the $x\,-\,y$ and $M\,-\,y$ diagrams show lines with constant $Q^2_{meas}$.

\subsection{Determination of the energy of the ISR photon}
In selecting DIS events the HERA experiments H1 and ZEUS place a cut on the difference between the total energy and longitudinal momentum 
\begin{eqnarray}
\delta = \sum (E\,-\,p_Z)     
\end{eqnarray}
observed in the detector which puts a limit on the energy of possible ISR. For a hermetic detector $\delta = 2E_{ebeam}$.  If the ISR photon escapes detection the requirement $\delta \ge \delta_{min}$ limits the photon energy: 
\begin{eqnarray}
E_{\gamma} \,=\, E_{ebeam} - \delta/2 \,<\,E_{ebeam} - \delta_{min}/2
\label{eq:fisr}
\end{eqnarray}
and therefore
\begin{eqnarray}
f_{\gamma} \, <\,  1 - \delta_{min}/(2 E_{beam}),
\label{eq:eisr} 
\end{eqnarray}
provided that measurement errors and other detector effects can be neglected. 

ISR in general affects the variables $x,y,Q^2$ differently when measured with the electron, DA or JB methods. For instance, in the presence of ISR the $x$ value measured with the electron method is always smaller than the true value, $x_{true}$, while the $x$ value obtained with the DA method is always larger than $x_{true}$. Turning the argument around, if one observes $x_{emeas} < x_{DAmeas}$ and $Q^2_{emeas} < Q^2_{DAmeas}$ then this is a strong indication for the emission of ISR, provided measurement errors can be neglected.
   
The energy carried by the ISR photon can be determined directly from a comparison of the variables found with the different methods. Assuming that measurement errors can be neglected one obtains\\

from the comparison of e and DA variables:
\begin{eqnarray}
f_{\gamma} & = & 1\,-\, \frac{E^{\prime}_{e\,emeas}}{E^{\prime}_{e\,DAmeas}}              \\
           & = & y_{emeas}(1\,-\,\frac{x_{emeas}}{x_{DAmeas}})   \nonumber  \\
                                                                 \nonumber  \\            & = & \frac{y_{emeas}\,-\,y_{DAmeas}}{1\,-\,y_{DAmeas}}       \nonumber  \\
           & = & 1\,-\, \frac{Q^2_{emeas}}{Q^2_{DAmeas}},                 \nonumber
\end{eqnarray}
from the comparison of e and JB variables:
\begin{eqnarray}
f_{\gamma} & = & y_{emeas}\,-\,y_{JBmeas}                  \\
           & = & (1\,-\,y_{JBmeas}) (1\,-\,\frac{Q^2_{JBmeas}}{Q^2_{emeas}}) \nonumber
\end{eqnarray}
and from the comparison of DA and JB variables:
\begin{eqnarray}
f_{\gamma} & = & 1\,-\,\frac{y_{JBmeas}}{y_{DAmeas}}        \\
           & = & (1\,-\,y_{JBmeas}) (1\,-\,\frac{x_{JBmeas}}{x_{DAmeas}})  \nonumber 
\end{eqnarray}

In the $\Sigma$ method the result is
\begin{eqnarray}
f_{\gamma} & = & 1\,-\, \frac{\sum_h(E_h\,-\,p_{Zh})\,+\,E^{\prime}_e(1\,-\,cos \theta^{\prime}_e)}{2E_{ebeam}}.
\end{eqnarray}
Of course, the most precise result for NC processes on $f_{\gamma}$ can be expected from a common fit to all variables measured as advocated in~\cite{Chaves}.

As mentioned before, for CC processes, the ISR energy must be determined from the direct observation of the photon. The information from the hadron side allows only to put a limit on the photon energy: $f_{\gamma} < (1\,-\,y_{JBmeas})$. 
 
\section{Acknowledgements}
I am grateful to Profs. D. Acosta and A. Caldwell, Dr. U. Katz and Prof. R. Klanner as well as to Drs. U. Bassler and G. Bernardi for helpful comments.

\newpage

\end{document}